# Anomalies in the thermal conductivity of honeycomb antiferromagnet $MnPS_3$

Jian Yan[1,2*], Hiromu Okamoto[2], Hiroki Yoshida[2], Hikaru Takeda[2], Xuan Luo[3], Yuping Sun[3,4,5], Jun-ichi Yamaura[2] and Minoru Yamashita[2*]

[1]*Institute for Advanced Study, Chengdu University, Chengdu, 610106, China*

[2] *The Institute for Solid State Physics, The University of Tokyo, Kashiwa, 277-8581, Japan*

[3] *Key Laboratory of Materials Physics, Institute of Solid State Physics, HFIPS, Chinese Academy of Sciences, Hefei, 230031, China*

[4] *Anhui Provincial Key Laboratory of Low-Energy Quantum Materials and Devices, High Magnetic Field Laboratory, HFIPS, Chinese Academy of Sciences, Hefei, 230031, China*

[5] *Collaborative Innovation Center of Advanced Microstructures, Nanjing University, Nanjing, 210093, China*

## Abstract

Intrinsic two-dimensional magnets serve as a good platform to explore collective, charge-neutral and low-energy excitations. Distinguishing the crucial role of them in experimental aspect remains a challenge for decades. Here, we study the thermal transport in honeycomb antiferromagnet $MnPS_3$ with $T_\mathrm{N} = 78$ K down to very low temperatures ($< 0.01T_\mathrm{N}$). At high temperatures ($> 0.1T_\mathrm{N}$), the field dependence of the thermal Hall conductivity exhibits a linear phonon Hall effect and a peak associated with the spin-flop transition due to a strong spin-lattice coupling, well reproducing the previous report (Phys. Rev. B 110, 165147 (2024)). Notably, below 2 K, we find that the field dependence of the thermal Hall conductivity exhibits sign reversals within the spin-flop phase, at which the field dependence of the longitudinal thermal conductivity also shows multiple valleys. We suggest that these anomalies are caused by the redistribution of Berry curvature in magnon bands, demonstrating the superior performance of the thermal Hall measurements to detect the Berry curvature distributions in magnetic insulators.

*Contact authors: yanjian@cdu.edu.cn and my@issp.u-tokyo.ac.jp

## Introduction

In recent years, the exploration of collective, charge-neutral, low-energy excitations in magnetic insulators has attracted considerable interest due to their potential for innovative approaches to manipulate and control thermal energy and information [1,2]. To study these charge-neutral excitations in magnetic insulators, a thermal Hall effect (THE), which is a thermal analog of electrical Hall effects, has been pointed out as a powerful tool [3-6]. Indeed, thermal Hall measurements have been adopted to study various charge-neutral excitations including magnons in ordered magnets [7-11], phonons [12-18], spinons in kagome antiferromagnets [19-22], and Majorana fermions in Kitaev candidate materials [23-28]. Although topological THEs by magnetic skyrmions have been studied in ferromagnetic and antiferromagnetic (AFM) three-dimensional magnets [9,10], the correspondence between the experiments and the theoretical calculations has yet to be clarified because of the complexity of the spin Hamiltonian of these three-dimensional materials.

The two-dimensional (2D) van der Waals magnets provide an ideal platform not only to understand the mechanism of low-dimensional magnetism [29-32] but also to study the topological magnon transports [33,34] because of its good two dimensionality. Among these 2D magnets, the family of transition metal thiophosphates $M\mathrm{PS}_3$ ($M$ = Mn, Ni, and Fe) with $M$ atoms forming a honeycomb lattice stacked by van der Waals interaction (Fig. 1(a)) has been pointed out to be suited to study the two-dimensional magnetism because these compounds are known to exhibit Heisenberg-type, XY or XXZ-type, and Ising-type antiferromagnetism for $M$ = Mn, Ni, and Fe, respectively [35-38]. In $\mathrm{MnPS}_3$, which has the monoclinic crystal structure with the space group $C2/m$, the magnetic moments of $\mathrm{Mn}^{2+}$ ions ($S = 5/2$) align antiferromagnetically along the direction tilted 8° from the $c^*$ axis to the $a$ axis below the Néel temperature of $T_\mathrm{N} = 78$ K due to the AFM spin interactions [39-44]. Due to this tilting, a spin-flop (SF) transition is induced by applying the magnetic field along the $c^*$ axis, as described in Fig. 1(b). This noncollinear magnetic order induced by the SF transition is pointed out to realize not only a change in the longitudinal thermal transport through its large spin-lattice coupling [45-47], but also a finite THE that change drastically at the transition point [34,48] even without the Dzyaloshinskii-Moriya interaction in this material [44]. Previous THE measurements of $\mathrm{MnPS}_3$ done above 20 K and below 9 T have succeeded in detecting finite thermal Hall conductivity $\kappa_{xy}$, from which both an exchange striction and a single-ion magnetostriction are discussed to play important roles in the THE of $\mathrm{MnPS}_3$ [49].

In this study, we extend the thermal-transport measurements of $\mathrm{MnPS}_3$ to lower temperatures ($< 0.01T_\mathrm{N}$) and higher magnetic fields. We find that the field dependence of the longitudinal thermal conductivity ($\kappa_{xx}$) below 50 K shows a large field suppression inside the SF phase, in addition to a sharp dip at the SF transition, showing a large contribution of the magnon in $\kappa_{xx}$. Notably, below 2 K, we find that the field dependence of the thermal Hall conductivity ($\kappa_{xy}$) shows multiple positive and negative peaks in the SF phase, which exhibits a sign reversal as lowering the temperature. Simultaneously, the field dependence of $\kappa_{xx}$ shows multiple valleys at the fields corresponding to the peaks in $\kappa_{xy}$. We find that these results are consistent with the theoretical calculations showing topological transitions of the magnon bands by the magnon band inversions [34] or multiple band crossings by the magnon-phonon hybridizations [48,49]. We suggest that these anomalies observed in the field dependence of $\kappa_{xx}$ and $\kappa_{xy}$ in the SF phase at lower temperatures are caused by a redistribution of Berry curvatures in the magnon bands.

## Materials and Methods

$MnPS_3$ single crystals used in this work were grown by the chemical vapor transport method using $I_2$ as the transport agent. Manganese powder (99.98%, Alfa Aesar), red phosphorus powder (99.999%, Alfa Aesar), and sulfur powder (99.999%, Alfa Aesar) with a mole ratio of 1:1:3.05, and 0.15 g of $I_2$ as transport agent were weighed and loaded into a silicon quartz tube in an argon atmosphere, which was sealed under a high vacuum. 5 wt% extra sulfur was added for diminishing the vacancy in sulfur sites. The sealed quartz tube was placed in a two-zone tube furnace with the hot side of approximately 680 C° and the cold side of 630 C°. The whole growth process lasts approximately 7 days and then slowly cooled down to room temperature at a rate of 100 C°/h. Finally, lots of hexagonal green $MnPS_3$ single crystal with the largest size of ~10 mm × 10 mm were obtained [50]. See Appendix A for the X-ray diffraction data and pictures of our samples.

The magnetization ($M$) of the $MnPS_3$ samples was measured using MPMS-XL5 (Quantum Design) above 2 K and a home-built Faraday-force magnetometer below 2 K by applying the magnetic field along the $c^*$ axis. The thermal-transport measurements were performed by the standard steady-state method by using one-heater-three-thermometers setup in a variable temperature insert for 2 K to 100 K and a dilution refrigerator below 2 K. The heat current $J_{\mathrm{Q}}$ and the magnetic field $B$ was applied parallel to the $a$ ($\parallel x$) and $c^*$ ($\parallel z$) axis of the sample, respectively [Fig. 1(c)]. Both the longitudinal ($\Delta T_x = T_{\mathrm{High}} - T_{\mathrm{L1}}$) and the transverse ($\Delta T_y = T_{\mathrm{L1}} - T_{\mathrm{L2}}$) thermal differences were simultaneously measured to determine $\kappa_{xx}$ and $\kappa_{xy}$ (see Appendix B for details).

## Results and Discussions

As shown in Fig. 1(d), the temperature dependence of the magnetic susceptibility ($\chi$) exhibits a broad maximum at around 100 K, which is followed by a sharp cusp at the Néel temperature $T_{\mathrm{N}} = 78$ K of $MnPS_3$ without discernible difference between the field cooling and the zero-field cooling, reproducing previous results [32,35,46,49]. The magnetic field dependence of the magnetization ($M$) at low temperatures shows a clear kink at the SF transition field ($B_{\mathrm{sf}} = 4.5$ T) that is determined by the peak of the field derivative of $M$ ($\mathrm{d}M/\mathrm{d}B$) as shown in Fig. 8 in Appendix C. This kink in the field dependence of $M$ at $B_{\mathrm{sf}}$ blurs with increasing temperatures. Above 50 K, only a linear field dependence of $M$ is observed.

Figure 1(e) shows the temperature dependence of $\kappa_{xx}$ under 0 T and 15 T ($B \parallel c^*$). As shown in Fig. 1(e), the temperature dependence of $\kappa_{xx}$ at both 0 T and 15 T exhibits a single peak at around 20 K, which is a general temperature dependence of a phonon thermal conductivity resulting from the competition between the increase of the phonon mean free path and the decrease of the phonon heat capacity as lowering the temperature [51]. The temperature derivative of $\kappa_{xx}$ displays a clear kink at $T_{\mathrm{N}}$, indicating a change in the phonon scatterings by the magnetic order through the strong spin-lattice coupling in $MnPS_3$, consistent with the drastic shift of Raman peak below $T_{\mathrm{N}}$ [46]. A clear suppression of $\kappa_{xx}$ by applying 15 T along the $c^*$ axis is observed below around 50 K due to a suppression of a magnetic component in $\kappa_{xx}$ as discussed below. Below 1 K, $\kappa_{xx}$ shows a temperature dependence of $\kappa_{xx} \propto T^{2.8}$ with the mean free path comparable with the thickness of the sample, showing a ballistic phonon conduction observed in a high-quality sample (see Fig. 9 in Appendix D).

Figure 2(a) shows the field dependence of the normalized magnetothermal conductivity defined as $\Delta\kappa_{xx}(B)/\kappa_{xx}(0) \equiv [\kappa_{xx}(B) - \kappa_{xx}(0)]/\kappa_{xx}(0)$ at selected temperatures. As shown in Fig. 2(a), $\Delta\kappa_{xx}(B)$ at 50 K shows a monotonical decrease up to 15 T. At lower temperature, $\Delta\kappa_{xx}(B)$ starts to show a sharp decrease above $B_{\mathrm{sf}}$ in addition to the monotonical decrease, concomitant with the clear indication of kink in the field dependence of $M$. We note that our data well reproduces the previous study reporting $\Delta\kappa_{xx}(B)$ above 9 K and below 9 T [49]. At 2 K, $\Delta\kappa_{xx}(B)$ shows a positive magnetothermal conductivity below $B_{\mathrm{sf}}$, followed by a dip at $B_{\mathrm{sf}}$ showing a clear correlation between $\Delta\kappa_{xx}(B)$ and the change of the magnetic state. Furthermore, $\Delta\kappa_{xx}(B)$ at 2 K shows a field suppression above 6 T, which is followed by an increase above 11 T.

This field dependence of $\Delta\kappa_{xx}(B)$ shows that $\kappa_{xx}$ of $MnPS_3$ contains both the phonon ($\kappa_{xx}^{\mathrm{ph}}$) and the magnetic ($\kappa_{xx}^{\mathrm{mag}}$) components. It is known that $\kappa_{xx}^{\mathrm{ph}}$ usually increases under a magnetic field because the magnetic field suppresses the spin fluctuations that scatter the phonons, which is observed as the positive magnetothermal conductivity below $B_{\mathrm{sf}}$ at 2 K (Fig. 2(a)) as well as the one observed above 100 K reported in the previous work [49]. A field suppression of $\kappa_{xx}^{\mathrm{ph}}$ is induced by additional scattering on phonons, which can be caused by a closing of a magnon gap or resonance scatterings [22,52]. The former can explain the dip in $\Delta\kappa_{xx}$ $(B)$ at $B_{\mathrm{sf}}$ and at 2 K; The Zeeman splitting of the degenerated magnon bands of the up and down spins at zero field induces the zero-energy crossing of the lower energy band at $B_{\mathrm{sf}}$. This closing of the magnon gap increases the low-energy magnons that scatter phonons through the magnon-phonon coupling [45-47]. On the other hand, the field suppression of $\Delta\kappa_{xx}$ $(B)$ above 6 T is inconsistent with either the closing of the magnon gap and the resonance scattering, because no further zero-energy touching of the magnon bands is expected in this material [34,48] and the magnetic field giving the minimum of $\Delta\kappa_{xx}$ $(B)$ in the resonance scattering should be lower than 6 T and should decrease with decreasing temperature [22] (1.5 T and 3.7 T at 2 K and 5 K, respectively for $g =$ 2 [53]) whereas the minimum of $\Delta\kappa_{xx}$ $(B)$ is observed at around 11 T both at 2 K and 5 K. Therefore, the field suppression of $\Delta\kappa_{xx}$ $(B)$ above 6 T should be attributed to a decrease of $\kappa_{xx}^{\mathrm{mag}}$ by the field increase of the magnon gap in the SF phase, indicating in turn a considerable contribution of $\kappa_{xx}^{\mathrm{mag}}$ in $\kappa_{xx}$ at low fields. The field increase of $\Delta\kappa_{xx}$ $(B)$ above 12 T observed below 5 K shows that the increase of $\kappa_{xx}^{\mathrm{ph}}$ by the field suppression effect on the magnon-phonon scatterings exceeds the decrease of $\kappa_{xx}^{\mathrm{mag}}$ by the increase of the magnon gap at higher fields.

Having established both the phonon and the magnon contributions in $\kappa_{xx}$, we now discuss the field dependence of $\kappa_{xy}$. As shown in Fig. 2(b), the field dependence of $\kappa_{xy}$ above 30 K shows a negative linear increase as $\kappa_{xy} = -R_0 B$. Below 20 K, the field dependence of $\kappa_{xy}$ starts to show peaks and valleys for $B > B_{\mathrm{sf}}$ in addition to the linear field dependence. This linear field dependence in $\kappa_{xy}(B)$ can be attributed to a phonon thermal Hall effect. Indeed, as shown in Fig. 2(c), the temperature dependence of $R_0$, which is estimated by a linear fit of the data above 12 T (the dash-dotted lines in Fig. 2(b)), well follows that of $\kappa_{xx}$ at 15 T, where $\kappa_{xx}^{\mathrm{ph}}$ becomes dominant. This linear scaling between the temperature dependence of $\kappa_{xy}$ and that of $\kappa_{xx}^{\mathrm{ph}}$ is known as a tell-tale signature of the phonon thermal Hall effect as reported in non-magnetic insulators [16,17]. Therefore, to investigate the details of the magnetic contribution in $\kappa_{xy}$ of $MnPS_3$, we subtract this linear component from $\kappa_{xy}$ data to estimate $\Delta\kappa_{xy} = \kappa_{xy} + R_0 B$.

The field dependence of the subtracted thermal Hall conductivity ($\Delta\kappa_{xy}$ $(B)$) is plotted in Fig. 3. As shown in Figs. 3(e) and 3(f), no discernible deviation from the linear field dependence is observed at 30 K and 50 K, showing the dominant phonon contribution in $\kappa_{xy}$. At 20 K, $\Delta\kappa_{xy}$ $(B)$ shows a positive peak at 6 T and a negative peak at 7.6 T, of which the magnetic field are denoted as $B_1$ and $B_2$, respectively. These peaks become larger with decreasing the temperature down to 5 K. Remarkably, at 2 K, the sign of these peaks of $\Delta\kappa_{xy}$ at $B_1$ and $B_2$ are reversed and a new negative peak appears at around 11 T (denoted as $B_3$). A weak positive peak might also appear at $B_{\mathrm{sf}}$. Given that a sign of magnon $\kappa_{xy}$ directly reflects the sign of the dominant Berry curvature of the magnon bands [3-5], the multiple sign reversals in the field dependence of $\Delta\kappa_{xy}$ indicate the reversals of the dominant Berry curvature of the magnon bands, suggesting multiple topological phase transitions within the SF phase [34]. We note that a similar sign reversal of $\kappa_{xy}$ observed in another antiferromagnet $Na_2Co_2TeO_6$ is also pointed out resulting from a topological transition of magnons by a SF transition [54,55]. We stress that, although the positive peak at $B_1$ is reported in the previous work [49], the additional peaks at $B_2$ and $B_3$ as well as its sign reversal at 2 K are new features revealed by our measurements done at lower temperatures.

To further confirm the origin of THE in $MnPS_3$, we focus on the ultralow temperature region down to $0.01T_{\mathrm{N}}$, where the phonon contributions is suppressed followed by the $T^3$ law. Figure 4(a) plots the field dependence of $\Delta\kappa_{xx}$ measured below 2 K in a different sample (#2). As shown in Fig. 4(a), the field dependence of $\Delta\kappa_{xx}$ at 2 K reproduces the data of sample #1, i.e. the dip at $B_{\mathrm{sf}}$, which is estimated to be almost the same with that in sample #1 (Fig. 10 in Appendix E), and the field suppression above 6 T followed by the increase above 8 T. Below 2 K, the dip at $B_{\mathrm{sf}}$ is split into two and a new minimum appears at $B_1$. The field of the first dip at 3.7 T may correspond to the onset of the SF transition and the second dip at 4.8 T to the peak of $\mathrm{d}M/\mathrm{d}B$. The anomaly at $B_1$ is also observed as the negative peak in $\kappa_{xy}$ of sample #2 at 2 K (Fig. 4(b)), reproducing the negative peak of $\kappa_{xy}$ in sample #1. These multiple valleys in the field dependence of $\kappa_{xx}$ is similar to that observed in the Kitaev candidate $\alpha$-$RuCl_3$ [56], which is pointed out to be caused by multiple field-induced phase transitions [57]. Although the field dependence of $\kappa_{xx}$ near the SF transition is in good agreement with the theoretical calculation of the phonon $\kappa_{xx}$ with a magnon-phonon hybridization [49], other features are not, indicating the presence of a magnetic contribution.

The field dependence of the magnetization of sample #2 was also measured at 0.5 K and 1 K up to 10 T (Fig. 11 in Appendix F). As shown in Fig. 11, except the kink at the SF transition, no obvious anomalies in $M(B)$ are observed within our resolution. This absence of anomaly in $M\ (B)$ at the fields at which valleys of $\kappa_{xx}$ or peaks in $\kappa_{xy}$ are observed shows the high sensitivity of $\kappa_{xx}$ and $\kappa_{xy}$ to slight changes in the magnon bands associated with magnetic transitions. The former detects changes of the low-lying magnon populations and its scatterings, and the latter changes in the distribution of Berry curvature. On the other hand, magnetization only reflects the response of spins, averaged in the whole Brillouin zone, to changes in the magnetic field. This different sensitivity to topological transitions is indeed predicted in the theoretical calculations [34].

Here, we discuss the origin of the multiple sign reversals of $\kappa_{xy}$ observed in the SF phase (Fig. 3(a)) based on the relation between the magnon $\kappa_{xy}$ and the Berry curvature. As described in Ref. [3,4], $\kappa_{xy}$ is given by a summation of the Berry curvature weighted by a function of Bose factor as

$$\frac{\kappa_{xy}}{T} = \frac{k_{\mathrm{B}}^2}{\hbar} \int_{\mathrm{BZ}} \Omega(\varepsilon) c_2[g(\varepsilon)] d\varepsilon, \qquad (1)$$

where $\Omega(\varepsilon)$ is the Berry curvature, $c_2(x) = (1+x)\left(\ln\frac{1+x}{x}\right)^2 - (\ln x)^2 - 2\mathrm{Li}_2(-x)$, $\mathrm{Li}_n(x)$ is a polylogarithm function $\mathrm{Li}_n(x)$, and $g(\varepsilon)$ is the Bose distribution function. This equation indicates that field dependence of a magnetic $\kappa_{xy}$ reflects the redistribution of the Berry curvature as a function of the field. Since this weighting factor $c_2(x)$ is a function that monotonically decreases with increasing energy, the Berry curvature of the lower energy band generally dominates $\kappa_{xy}$ for magnon bands with the opposite Berry curvatures in the upper and the lower energy bands [58]. Good agreements between the theoretical estimations of $\kappa_{xy}$ based on Eq. (1) and the experiments have been observed in various materials including the ferromagnetic skyrmion material [9] and the antiferromagnetic kagome [20].

There are several possible origins that give rise to a sign change of $\kappa_{xy}$. One possible origin is a topological phase transition in the AFM phase. Indeed, it is pointed out in Ref. [34] that the AFM spins on the honeycomb lattice under the magnetic field exhibit two topological phase transitions within the SF phase by the inversion of the magnon bands near the Brillouin-zone edge, in addition to the magnetic phase transitions into the SF phase. These topological phase transitions are manifested by the sign reversal of $\kappa_{xy}$, which are consistent with the multiple sign reversals observed in the field dependence of $\kappa_{xy}$ (Fig. 3). Such a topological phase transition can also be caused by a hybridization between the magnon and the phonon bands, in which a hybridization between the magnons and the phonons by a magnon-phonon coupling is pointed out to make band crossing points with Berry curvatures [48,49]. These band crossings, which may occur in the SF phase under the magnetic field, could also give rise to a sign reversal of $\kappa_{xy}$ due to a field-induced redistribution of the Berry curvature.

On the other hand, the sign reversal of $\kappa_{xy}$ at $B_1$ and $B_2$ from 5 K to 2 K is rather unexpected in Ref. [34]. Also, the critical fields and the temperature showing the sign reversal of $\kappa_{xy}$ predicted by Ref. [34] are much higher than the observed values of our experiment. These differences could result from a difference of the spin interaction parameters used in the calculations [34] from the actual ones in our samples. Indeed, the neutron scattering experiments [41,43] have shown the presence of further-neighbor interactions that are not included in the calculations [34], which would easily change the energy scale of the magnon band structure. Moreover, the SF field obtained in the calculation (~3.6 T) is smaller than that observed in our samples (4.5 T), implying a difference in the spin Hamiltonian. Therefore, it remains an important future work to confirm the observed sign reversals of $\kappa_{xy}$ in the calculations including the further-neighbor interactions.

The second possibility is that $\kappa_{xy}$ consists of two different origins (such as magnons and phonons) with opposite sign of $\kappa_{xy}$. In this case, the sign of $\kappa_{xy}$ will change depending on which component becomes dominant. Indeed, the sign reversals of $\kappa_{xy}$ at 5 K (Fig. 2(b)) are observed only near at $B_1$ where the positive magnetic component exceeds the negative linear background from the phonon contribution. As shown in Fig. 3, this linear phonon background needs to be subtracted to show the sign reversals of the magnetic component $\Delta\kappa_{xy}$. However, our $\kappa_{xy}$ measurements at 2 K (Fig. 2(b)) clearly demonstrate the sign reversals of $\kappa_{xy}$ without subtracting the linear background, excluding this two-component scenario as the origin of the sign reversals of $\kappa_{xy}$ in the SF phase of $MnPS_3$.

Another possibility is a contribution of the Berry curvature of the higher-energy bands at high temperatures. At a high temperature comparable to the spin interaction energy, a sign of $\kappa_{xy}$ is not simply determined by the Chern number of the lowest magnon band as shown by the calculations done in a ferromagnetic kagome [58,59]. In these cases, the contribution of the Berry curvature from higher-energy bands becomes larger at higher temperatures whereas those in lower-energy bands smear out by a thermal broadening effect, giving rise to a sign change of $\kappa_{xy}$ as a function of temperature or field at high temperatures. In fact, in the calculations considering a hybridization between the magnon and phonon bands [48,49], sign reversals of $\kappa_{xy}$ of $MnPS_3$ are shown to appear at around $B_{\mathrm{sf}}$ at higher temperatures due to the anisotropy of multiple band crossing points within the narrow energy and momentum range. On the other hand, these magnon-phonon couplings induce the sign reversals near $B_{\mathrm{sf}}$, not in the SF phase as observed in our measurements (especially at around 9 T in Fig. 3(a)). Moreover, the magnitude of $\kappa_{xy}$ calculated by this hybridization is orders of magnitude smaller than that observed by our measurements. Therefore, it also remains a challenge for future research to try to reproduce $\kappa_{xy}$ observed in $MnPS_3$ by introducing a magnon-phonon hybridization.

In summary, we have studied the thermal-transport properties in $MnPS_3$ down to very low temperatures ($< 0.01T_N$) and find that the field dependence of the thermal Hall conductivity exhibits multiple sign reversals above the SF transition field. Further, we find the field dependence of the longitudinal thermal conductivity also shows multiple valleys at the lowest temperature. We suggest that these anomalies are caused by the redistribution of Berry curvature in magnon bands, demonstrating the superior performance of the thermal Hall measurements to study the Berry curvature distributions in magnetic insulators.

## Acknowledgments

We acknowledge Hyun-Yong Lee, Je-Geun Park, and Heejun Yang for insightful discussions. J.Y. was supported by Grant-in-Aid for JSPS Fellows. J.Y. and M.Y. thank the support by JSPS KAKENHI Grant Numbers JP22KF0111 and JP23H01116. X.L. and Y.P.S. was supported by the National Key R&D Program of China (Grants No. 2023YFA1607402 and No. 2021YFA1600201), the National Natural Science Foundation of China (Grants No. 12274412, No. 12204487 and No. U2032215), the CASHIPS Director's Fund (Grant No. YZJJ2023QN34), and the Systematic Fundamental Research Program Leveraging Major Scientific and Technological Infrastructure, Chinese Academy of Sciences, under Contract No. JZHKYPT-437 2021-08.

## Data availability

All the data supporting this study is available from the corresponding authors upon reasonable request.

Figures

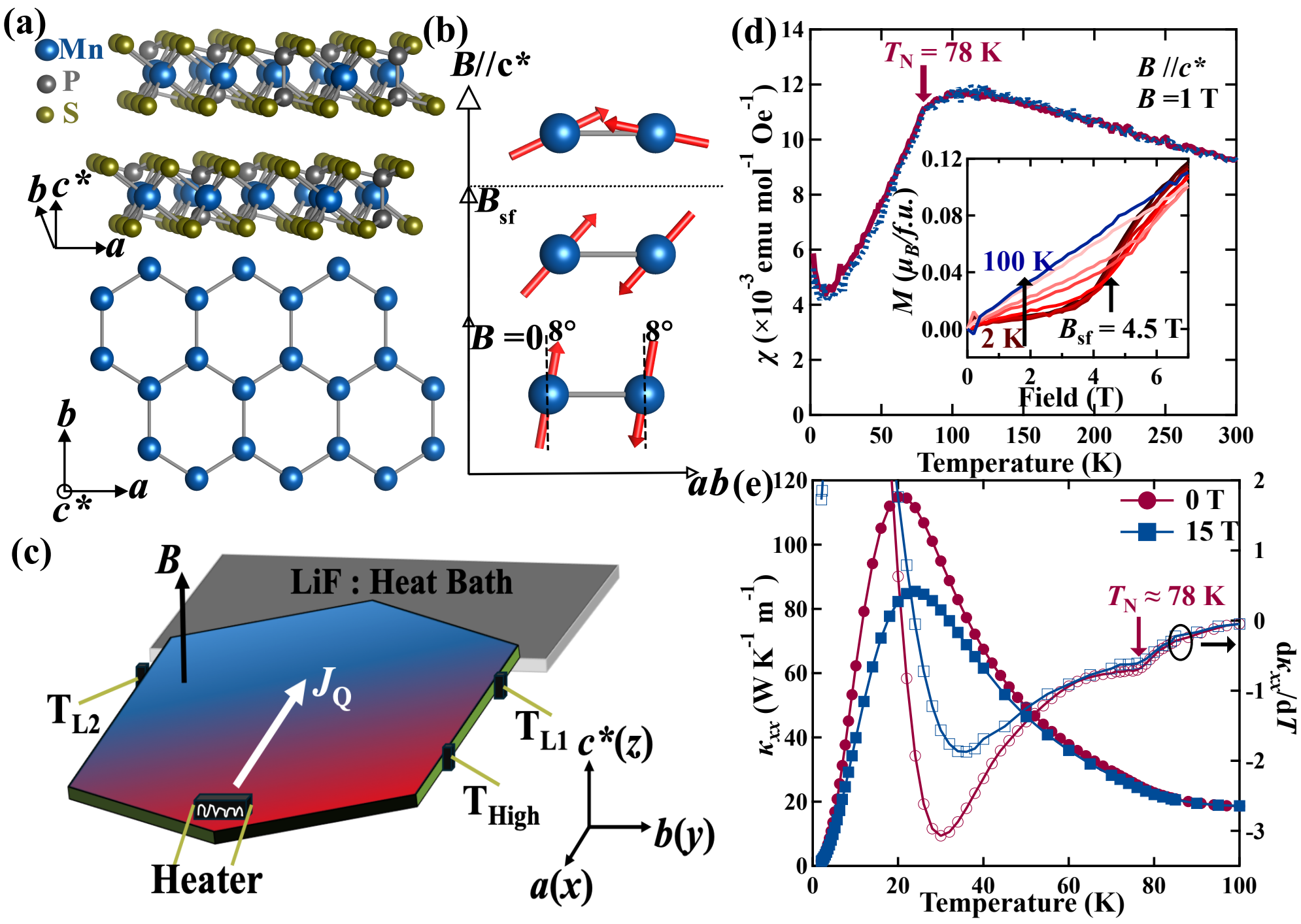

FIG. 1. (a) The side view of $MnPS_3$ crystal structure with van der Waals gap (top) and the top view of the $Mn^{2+}$ honeycomb layer (bottom). (b) Schematic of the spin configurations across the spin-flop transition field $B_{\mathrm{sf}}$. (c) The thermal-transport setup for measuring the longitudinal $(T_{\mathrm{High}} - T_{\mathrm{L1}})$ and the transverse $(T_{\mathrm{L1}} - T_{\mathrm{L2}})$ thermal gradients simultaneously as a function of the applied heat current $J_{\mathrm{Q}}$. (d) Temperature dependence of the magnetic susceptibility ($\chi$) measured at 1 T ($B \parallel c^*$) after zero-field cooling (ZFC, the dashed line) and field cooling (FC, the solid line). The inside shows the field dependence of the magnetization ($M$) at different temperatures. The spin-flop field $B_{\mathrm{sf}} = 4.5$ T is determined by the peak in the field derivative of $M$ as shown in Appendix C. (e) The temperature dependence of the longitudinal thermal conductivity ($\kappa_{xx}$) at 0 T (circles) and 15 T (squares). The temperature derivative of $\kappa_{xx}$ is shown on the right axis.

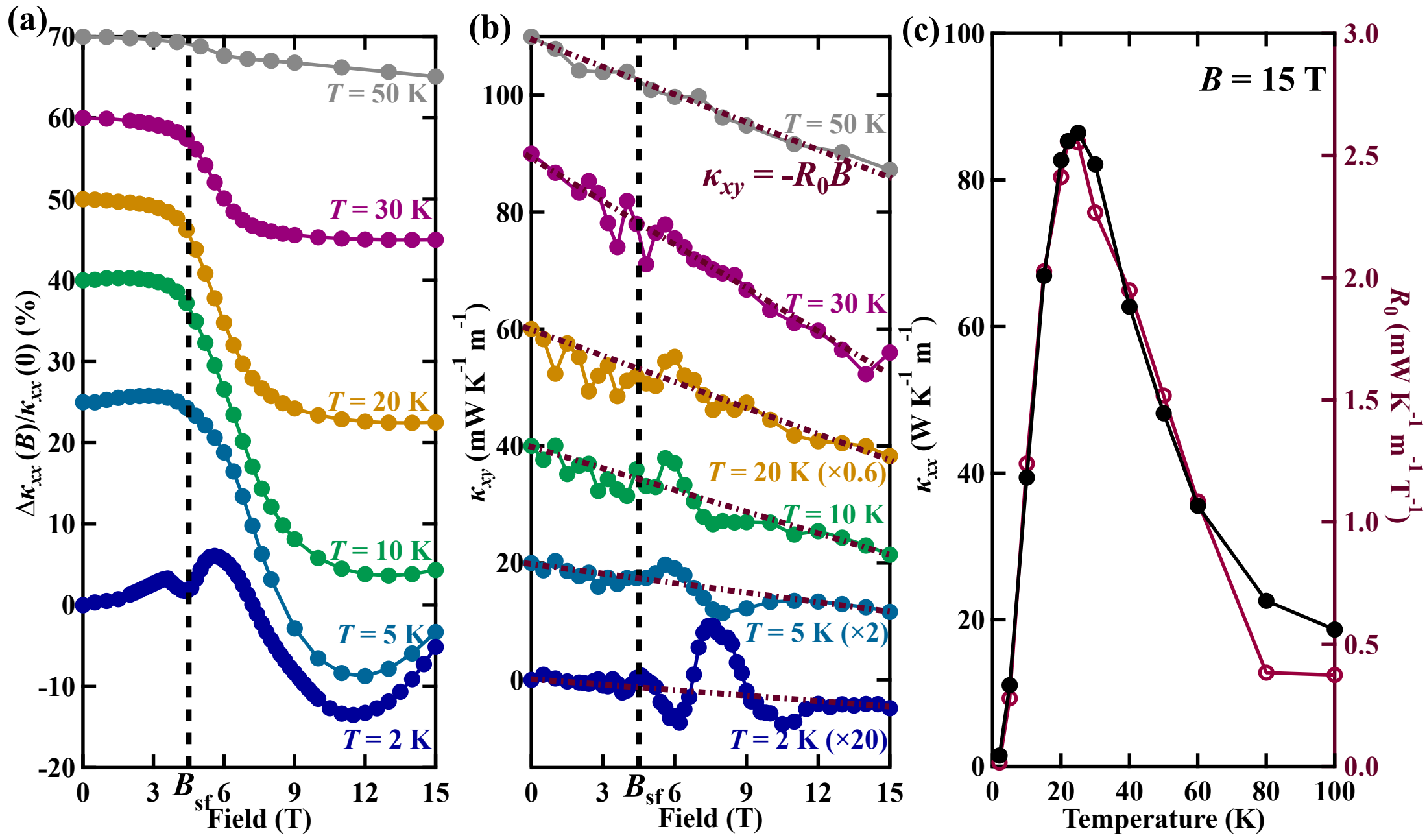

FIG. 2. (a,b) The field dependence of $\Delta\kappa_{xx}(B)/\kappa_{xx}(0)$ (a) and $\kappa_{xy}$ (b) at different temperatures. The vertical dashed-line shows $B_{\rm sf}$. The data are vertically shifted for clarity. The dash-dotted lines in (b) show the linear fit of the data to determine the slope of $R_0$. (c) The temperature dependence of $\kappa_{xx}$ at 15 T (left) and the that of the slope $R_0$ determined by the linear fit of the field dependence of $\kappa_{xy}$ (right).

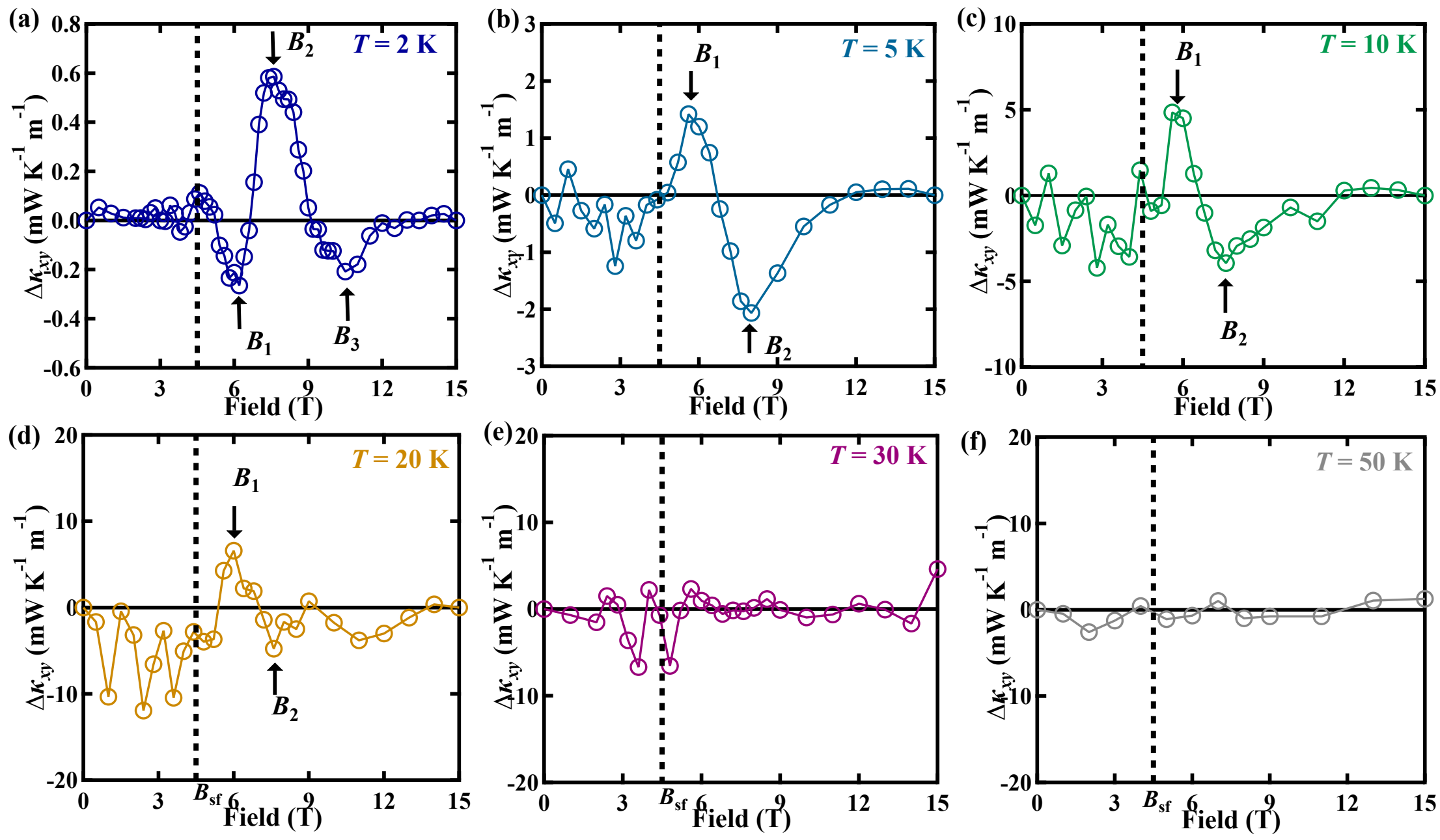

FIG. 3. (a–f) The field dependence of the subtracted $\Delta\kappa_{xy} = \kappa_{xy} + R_0 B$ at 2 K to 50 K. The vertical dashed line indicates $B_{\mathrm{sf}}$.

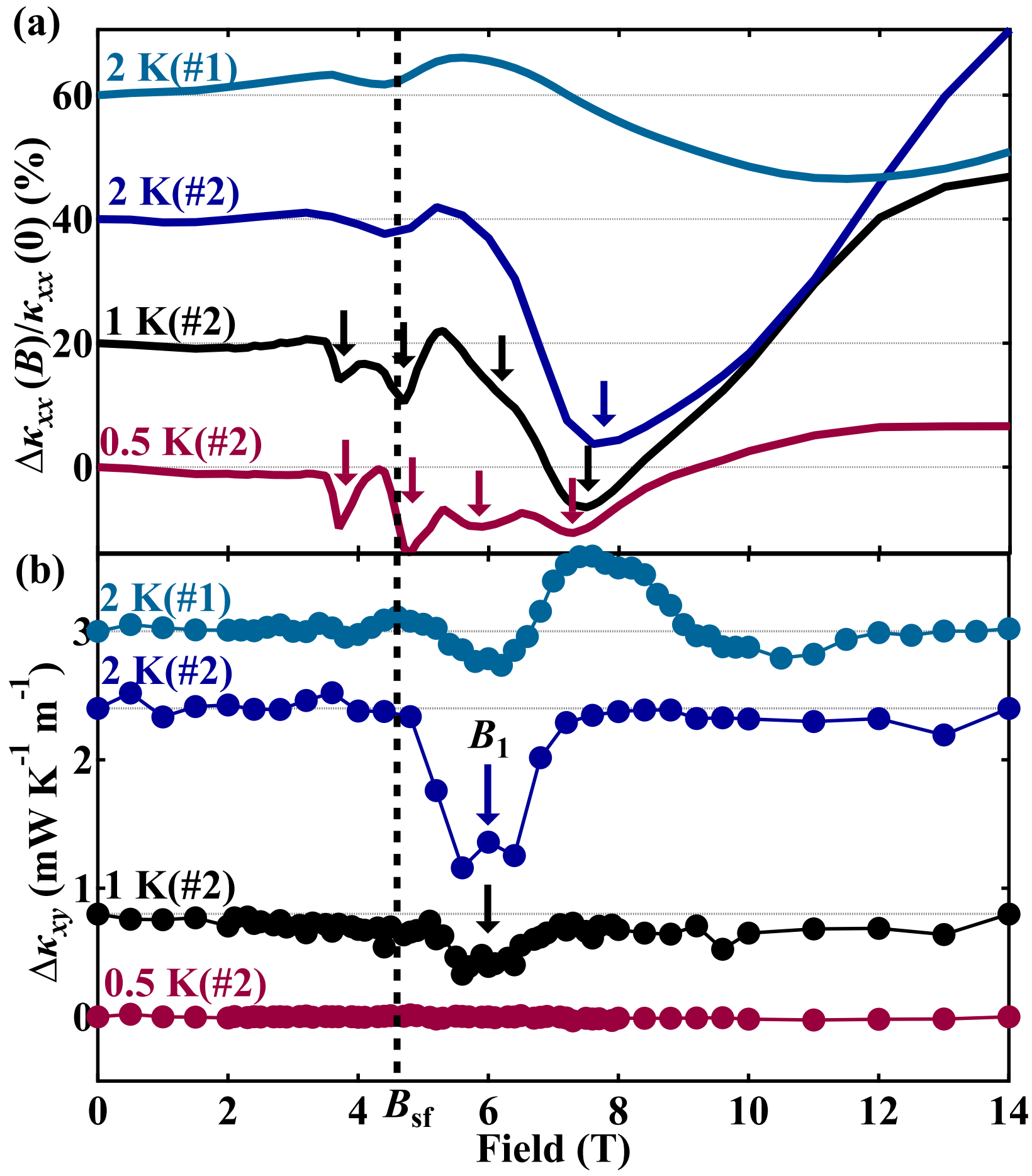

FIG. 4 (a, b) The field dependence of $\Delta\kappa_{xx}(B)/\kappa_{xx}(0)$ (a) and $\Delta\kappa_{xy}$ (b) measured in sample #2 at 0.5 K, 1.0 K, and 2.0 K. The data of sample #1 at 2 K in Fig. 2 is also shown for reference. The data is vertically shifted for clarity. The vertical dashed line indicates $B_{\mathrm{sf}}$.

## Appendix A: X-ray diffraction and single crystal pictures

The X-ray diffraction data clearly show $(00l)$ peaks, indicating that the surface of the crystals is $ab$-plane [Fig. 5(a)]. We exfoliated $MnPS_3$ with Scotch tape and chose two hexagonal shape samples with different thickness for the thermal-transport measurements. The sizes of two samples are about 3 mm × 3 mm × 0.0065 mm [#1, Fig. 5(b)] and 4 mm × 4 mm × 0.018 mm [#2, Fig. 5(c)]. We note that there is a large ambiguity in the estimation of the thickness of sample #1 because of the thinness, which results in an ambiguity in the estimation of the magnitude of $\kappa_{xx}$ of sample #1 up to around 30%.

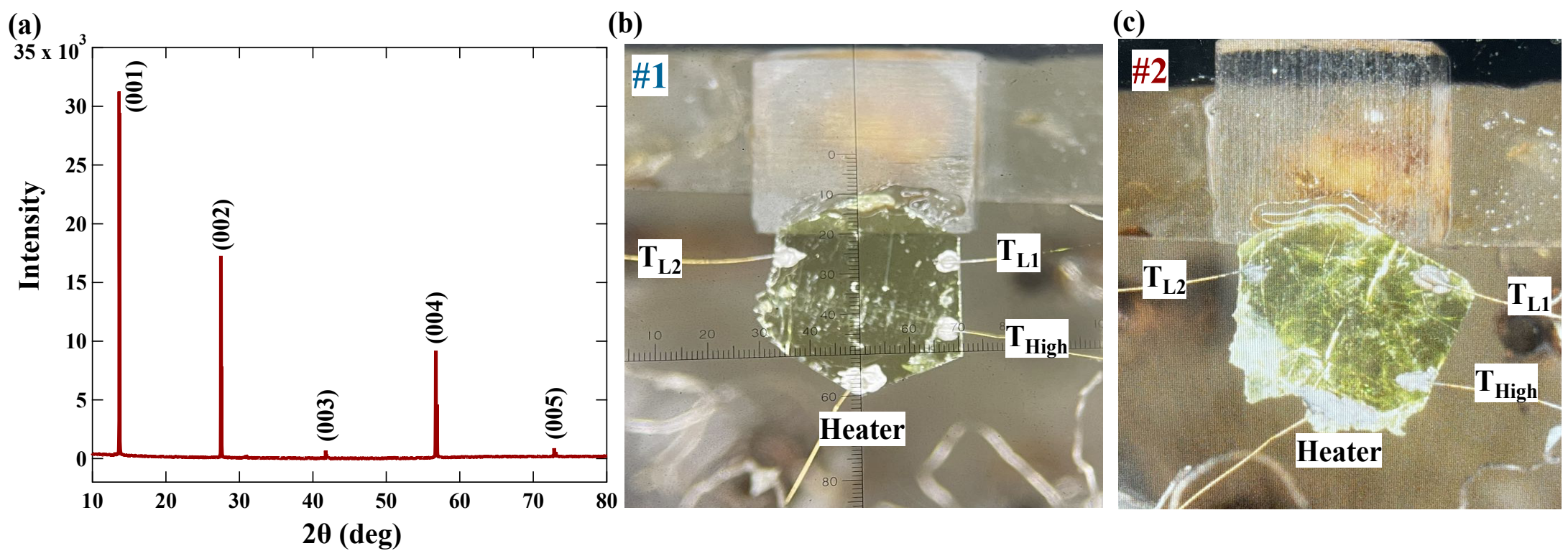

FIG. 5. (a) X-ray diffraction data of the $MnPS_3$ single crystal. (b,c) Pictures of the thermal-transport setup for sample #1 (b) and sample #2 (c).

## Appendix B: The setup of the thermal-transport measurements

For our thermal-transport measurements, One heater and three thermometers (Cernox™ (CX-1050) thermometers for measurements above 2 K and $RuO_2$ thermometers below 2 K) were attached to the sample via gold wires so that both the longitudinal ($\Delta T_x = T_{\mathrm{High}} - T_{\mathrm{L1}}$) and the transverse ($\Delta T_y = T_{\mathrm{L1}} - T_{\mathrm{L2}}$) thermal differences can be measured simultaneously. These resistive thermometers have the advantage of the high sensitivity at low temperatures. All thermometers were carefully calibrated in magnetic fields by a calibrated thermometer placed outside the magnetic field. The heat current $J_{\mathrm{Q}} = Q/wt$ was applied parallel to the $a$–$b$ ($x$–$y$) plane of the sample and the magnetic field was applied along the $c^*$ ($z$) axis, where $Q$ is the heater power, $w$ the width of the sample between the thermal contacts to read $T_{\mathrm{L1}}$ and $T_{\mathrm{L2}}$, and $t$ the thickness of the sample. To cancel the longitudinal component in $\Delta T_y$ due to a misalignment of the thermal contacts, $\Delta T_y$ was asymmetrized with respect to the field direction as $\Delta T_y^{\mathrm{asym}} = [\Delta T_y(+B) - \Delta T_y(-B)]/2$.

The longitudinal thermal conductivity $\kappa_{xx}$ and thermal Hall conductivity $\kappa_{xy}$ is derived by

$$\begin{pmatrix} Q/wt \\ 0 \end{pmatrix} = \begin{pmatrix} \kappa_{xx} & \kappa_{xy} \\ -\kappa_{xy} & \kappa_{xx} \end{pmatrix} \begin{pmatrix} \Delta T_x/L \\ \Delta T_y^{\mathrm{asym}}/w \end{pmatrix}, \tag{S1}$$

where $L$ is the length between $T_{\mathrm{High}}$ and $T_{\mathrm{L1}}$ [60]. In this estimation, the anisotropy in the $a$–$b$ plane due to the monoclinic crystal structure of this sample is ignored (i. e. $\kappa_{xx} = \kappa_{yy}$) because the distortion from a perfect hexagonal lattice is very small; the $b$ axis length is close to $\sqrt{3}a$ expected for the hexagonal lattice [39]. The small magnetic anisotropy in the $a$–$b$ plane has also been shown in the neutron scattering experiments [43].

The direction of the heat current was confirmed to be along the $a$ axis for sample #2 via the XRD measurements done after the thermal-transport measurements. Unfortunately, the direction of the $a$ axis of sample #1 could not be determined because the sample shattered into pieces when it was removed from the thermal-transport measurement cell after the measurements. Assuming the relationship between the hexagonal sample shape and the crystal axis remains the same with sample #2, it is likely that the heat current was also applied along the $a$ axis in sample #1.

According to Eq. (S1), $\Delta T_y^{\mathrm{asym}}$ is expected to linearly depend on $J_{\mathrm{Q}}$. We checked this linear relation by measuring the field dependence of $\Delta T_y^{\mathrm{asym}}$ at different $J_{\mathrm{Q}}$. As shown in Fig. 6, $\Delta T_y^{\mathrm{asym}}$ increases linearly with $J_{\mathrm{Q}}$ (the inset of Fig. 6(a)) and the field dependence of $\kappa_{xy}$ determined by $\Delta T_y^{\mathrm{asym}}$ measured with the different $J_{\mathrm{Q}}$ shows a good match, confirming the validity of our thermal Hall measurements.

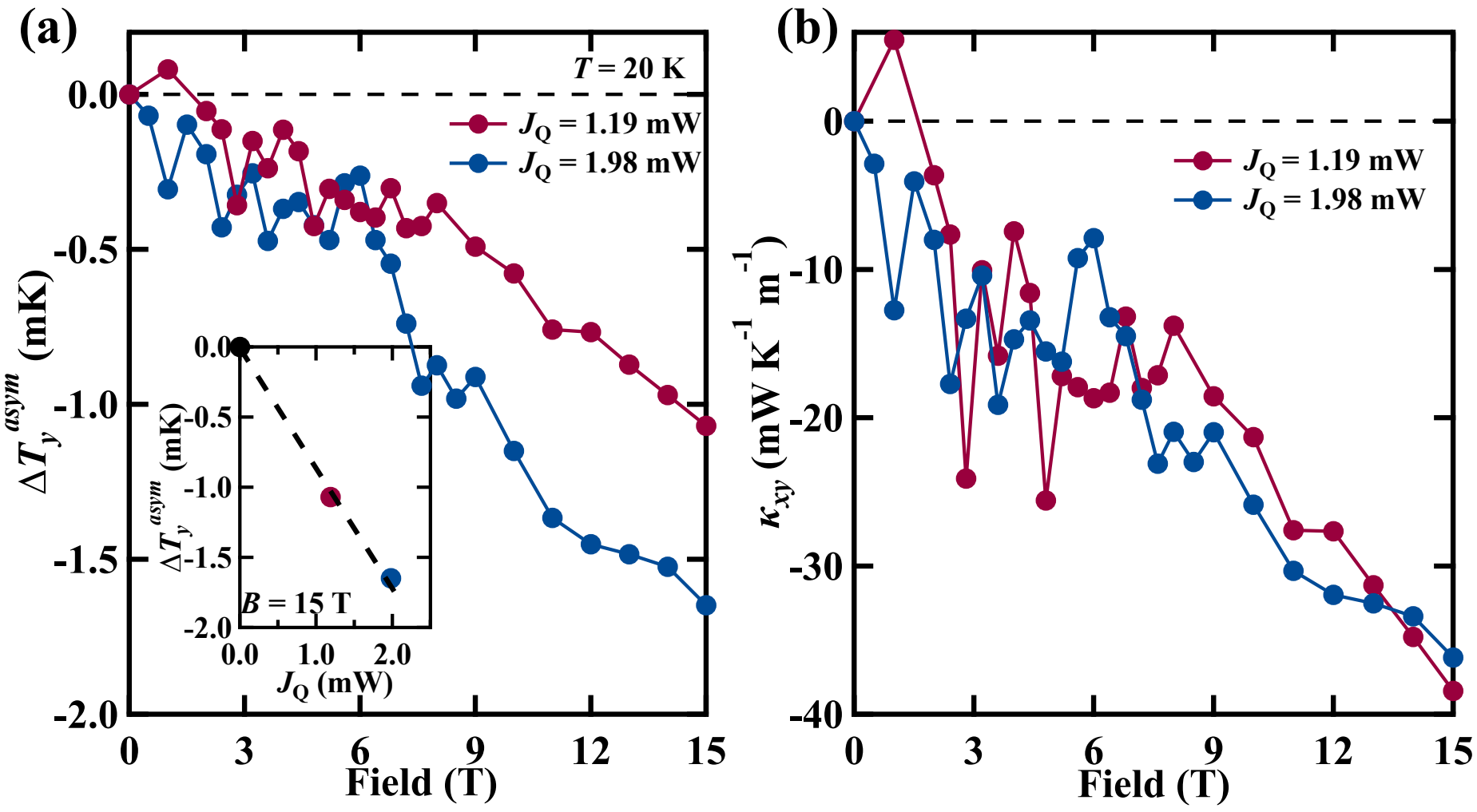

FIG. 6. (a) The field dependence of $\Delta T_y^{\mathrm{asym}}$ measured by applying different $J_{\mathrm{Q}}$. The inset shows $\Delta T_y^{\mathrm{asym}}$ measured at 15 T as a function of $J_{\mathrm{Q}}$. (f) The field dependence of $\kappa_{xy}$ determined from $\Delta T_y^{\mathrm{asym}}$ shown in (a).

Figure 7 shows the raw data of the temperatures during the measurements at 20 K shown in Fig. 6. The field dependence of $\Delta T_y^{\mathrm{asym}}$ is obtained by antisymmetrizing the data averaged for about 6 minutes before changing the heater status. The standard error of the data is typically about 0.020 mK, which is about the same size as the symbols in Fig. 6 (a).

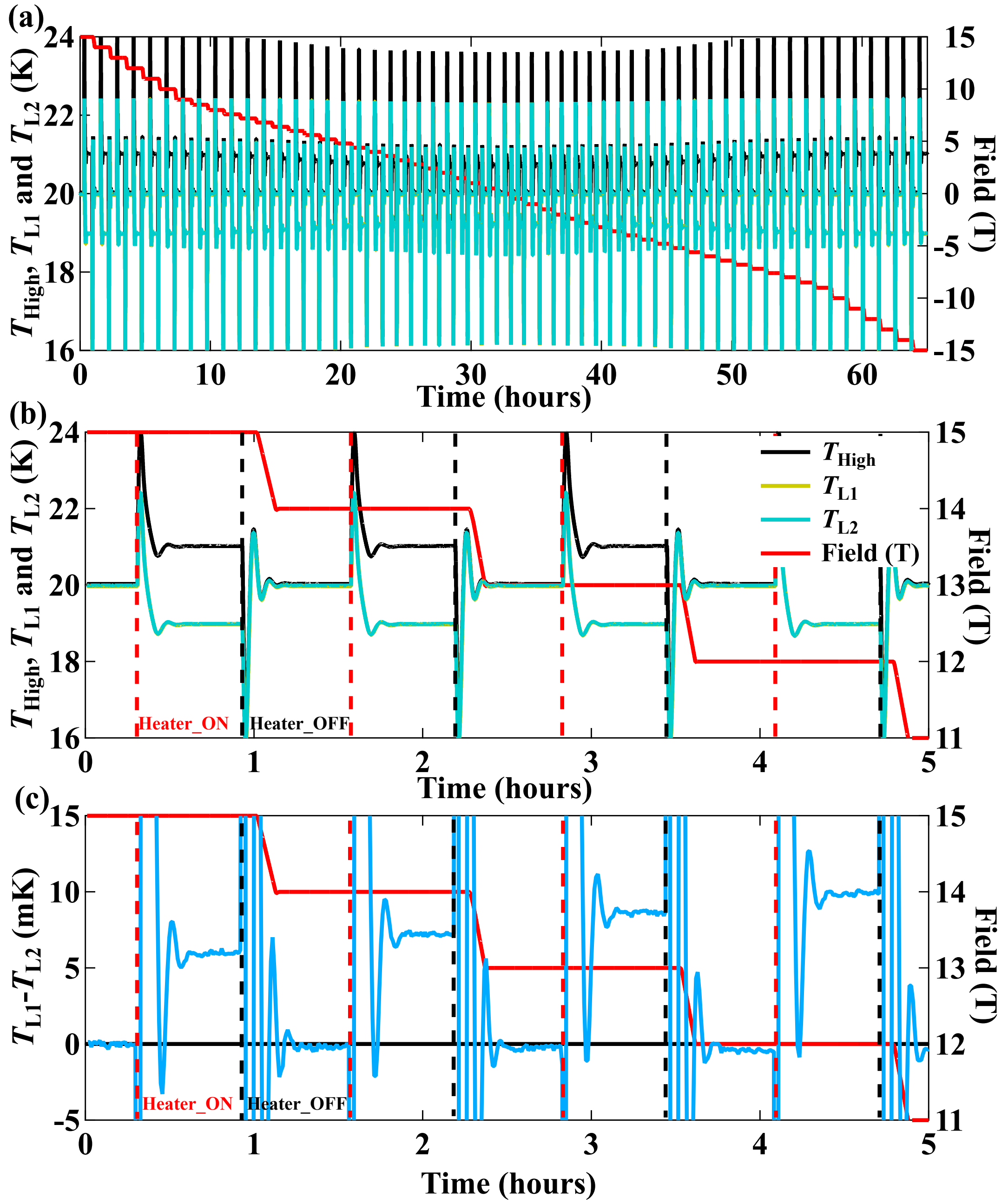

FIG. 7. (a) A raw time chart of the sample temperatures ($T_{\mathrm{High}}$ black line, $T_{\mathrm{L1}}$ yellow line, and $T_{\mathrm{L2}}$ cyan line) during the field sweep (red line, shown in the right axis) measurements at 20 K. (b) An enlarged figure of the first five hours data. The heater of 1.98 mW was turned on and off

at the time indicated by the red and black dashed line, respectively. The average temperature of $T_{\mathrm{High}}$ and $T_{\mathrm{L1}}$ was kept constant during the measurements. (c) A time chart of the transverse temperature difference $T_{\mathrm{L1}} - T_{\mathrm{L2}}$ (blue line) obtained from the data shown in (b).

**Appendix C: Determination of $B_{\mathrm{sf}}$.**

The spin-flop transition field ($B_{\mathrm{sf}}$) is determined by the peak of the field derivative of the magnetization ($\mathrm{d}M/\mathrm{d}B$) as shown in Fig. 8.

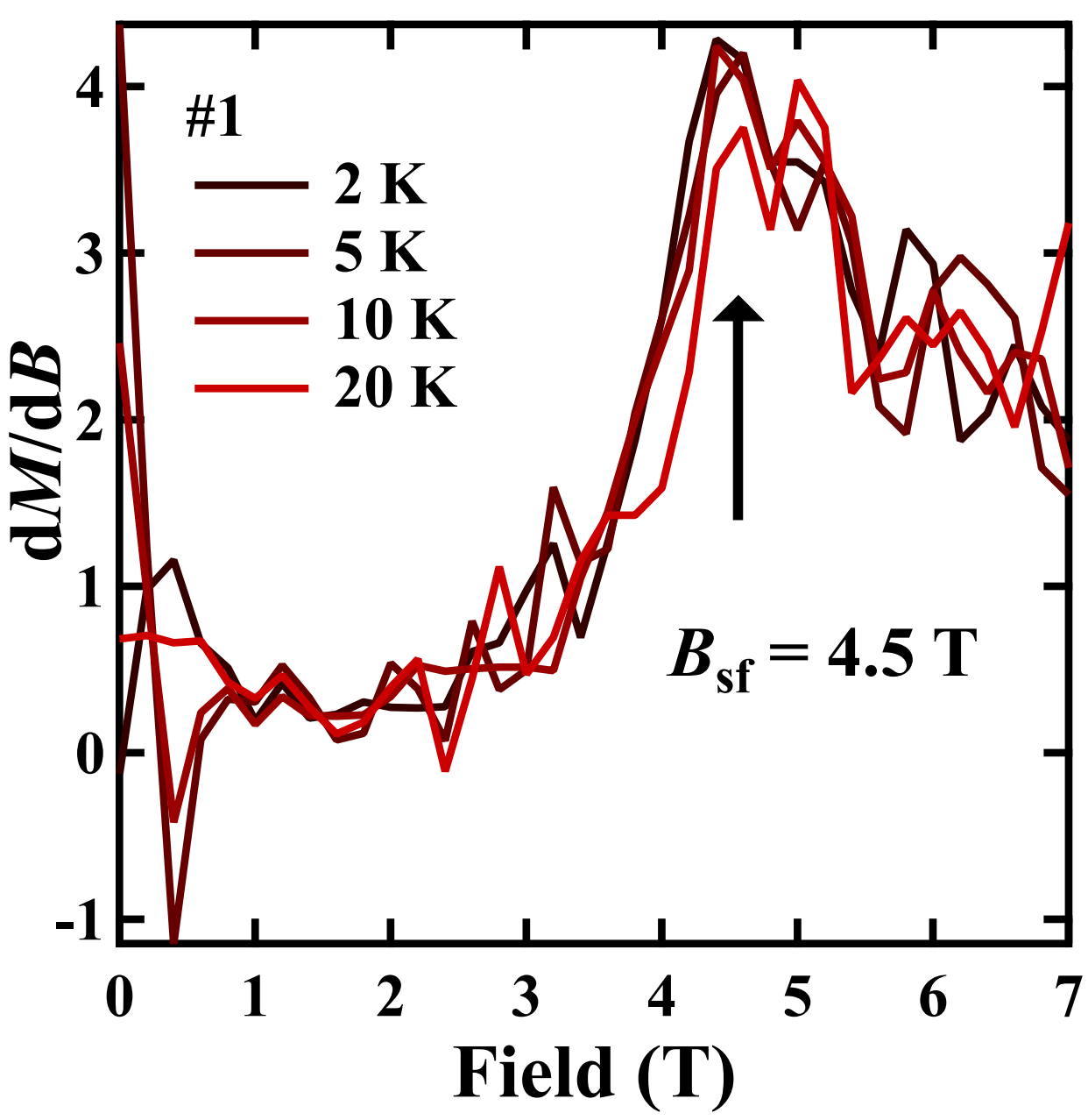

FIG. 8. the field derivative of $M$ ($\mathrm{d}M/\mathrm{d}B$) at selected temperatures.

**Appendix D: Estimation of the phonon mean free path**

Fig. 9(a) shows the temperature dependence of $\kappa_{xx}$ below 3 K. The power law fitting of the data ($\kappa_{xx} \propto T^n$) below 1 K gives the exponent of temperature dependence $n = 2.8$, which is close to that of boundary-limited phonons ($\kappa_{xx} \propto T^3$). In fact, the long mean free path of phonons comparable to the sample size is obtained as following. The thermal conductivity of phonons is given by

$$\kappa_{xx}^{\mathrm{ph}} = \frac{1}{3} C_{\mathrm{ph}} v_{\mathrm{ph}} l_{\mathrm{ph}}, \tag{S2}$$

where $C_{\mathrm{ph}}$, $v_{\mathrm{ph}}$, and $l_{\mathrm{ph}}$ is the heat capacity, the sound velocity, and the mean free path of phonons, respectively. We estimate $C_{\mathrm{ph}}$ by a $T^3$ fitting to the heat capacity data [61] as $C_{\mathrm{ph}}/T^3 = 5.7 \times 10^{-4}$ J mol$^{-1}$ K$^{-4}$ and the molar volume of $6.23 \times 10^{-5}$ m$^3$ mol$^{-1}$. The sound velocity $v_{\mathrm{ph}}$ is estimated from the phonon dispersion of $MnPS_3$ [49] as 3454 m/s. Assuming the dominant phonon contribution in $\kappa_{xx}$ below 1 K, the temperature dependence of the phonon mean free path (Fig. 9(b)) is obtained from the temperature dependence of $\kappa_{xx}$ (Fig. 9(a)). As shown in Fig. 9(b), $l_{\mathrm{ph}}$ exceeds the thickness of the sample below 3 K, demonstrating the ballistic phonon conduction observed in high-quality samples. The increase of $l_{\mathrm{ph}}$ beyond the thickness of the sample at lower temperature could be due to specular scatterings that is known to enhances $l_{\mathrm{ph}}$ beyond the sample thickness, or an additional magnon contribution in $\kappa_{xx}$.

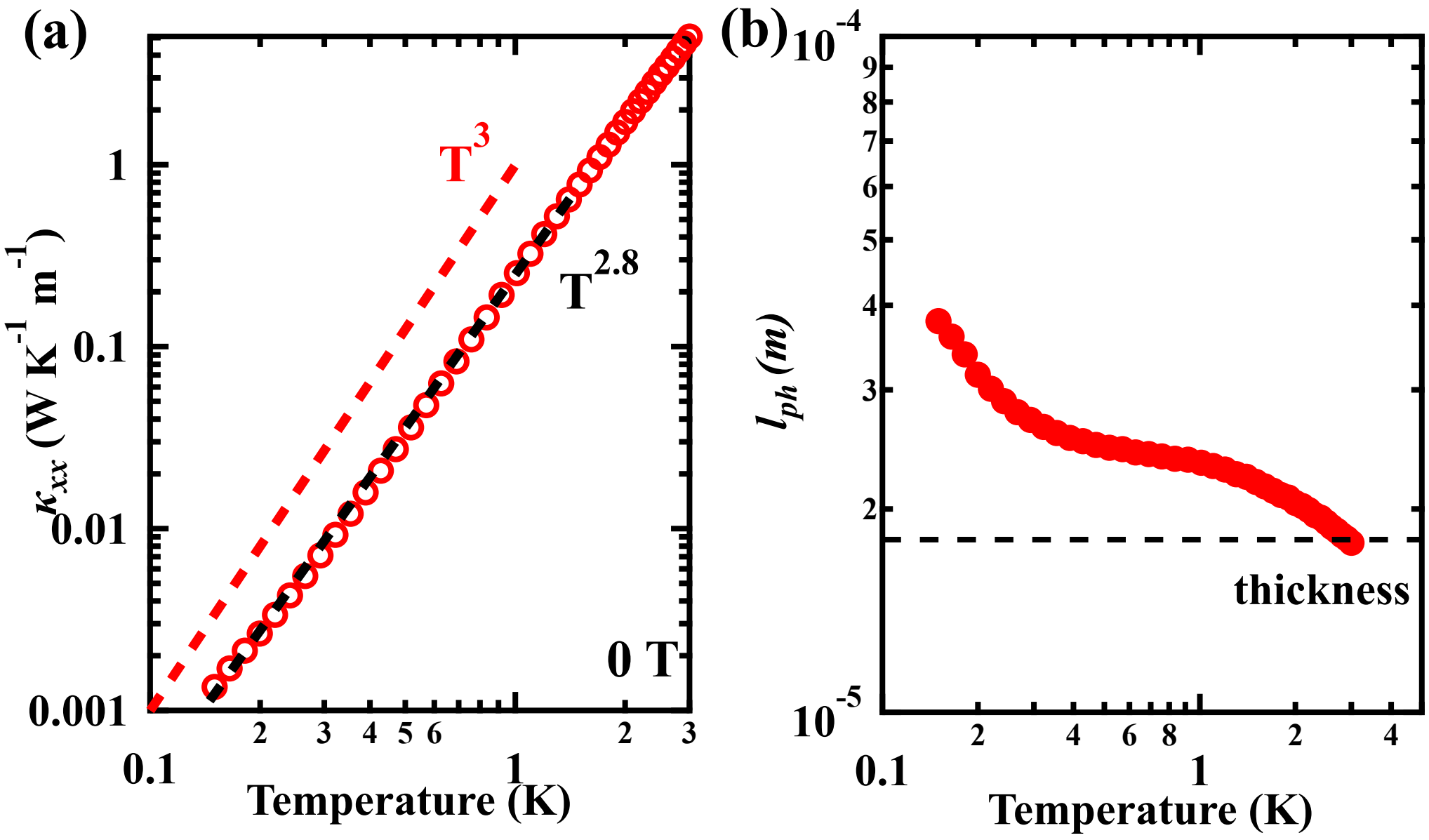

FIG. 9. (a) A log-log plot of the temperature dependence of $\kappa_{xx}$ of sample #2 measured in the dilution refrigerator. (b) The temperature dependence of the phonon mean free path at 0 T.

**Appendix E: The sample dependence of the magnetization and the thermal conductivity**

We checked the reproducibility of our data by measuring two $MnPS_3$ samples (#1 and #2). As shown in Fig. 10, the temperature dependence of the magnetic susceptibility (Fig. 10(a)), the field dependence of the magnetization (Fig. 10(b)), and the temperature dependence of the longitudinal thermal conductivity (Fig. 10(c)) of the two samples are very similar, showing the good reproducibility.

On the other hand, differences in the field dependence of $\kappa_{xx}$ and $\kappa_{xy}$ at 2 K are observed as shown in Fig. 4; the field dependence of $\Delta\kappa_{xx}$ of sample #2 at 2 K shows a larger minimum at a lower field of 8 T than that in sample #1, and the positive peak of $\Delta\kappa_{xy}$ of sample #1 at $B_2$ is not observed in sample #2.

One possible reason for this sample dependence is the difference in the sample thickness; the sample #2 ($t = 0.018$ mm) is about three times thicker than that of sample #1 ($t = 0.0065$ mm). Therefore, $l_{\rm ph}$ of sample #2 can be longer than that of sample #1, as shown by the larger $\kappa_{xx}$ of sample #2 (Fig. 10(c)). This difference in the phonon mean free path would result in a different relative magnitude of $\kappa_{xx}^{\rm mag}$ to $\kappa_{xx}^{\rm ph}$, and the different field dependence of $\kappa_{xx}$ due to a different field dependence of $\kappa_{xx}^{\rm mag}$ and $\kappa_{xx}^{\rm ph}$.

As shown in Fig. 10 (a), the magnetic susceptibility of sample #2 shows a larger Curie contribution than that in sample #1 below around 15 K, implying that sample #2 contains more magnetic impurities than that in sample #1. This difference in the amount of the magnetic impurities might also affect the field dependence of $\kappa_{xx}$ and $\kappa_{xy}$. It should be noted that the magnitude of $\kappa_{xx}$ of this compound is not a good measure of the sample quality. As shown in Fig. 10 (c), the clear phonon peak is observed in the temperature dependence of $\kappa_{xx}$ at around 20 K in both samples, showing that $\ell_{\rm ph}$ becomes comparable to the sample size, much longer than the impurity size. Thus, $\kappa_{xx}^{\rm ph}$ is no longer suppressed by these impurities.

In addition, this sample dependence might be brought by a different degree of distortions within the sample (e.g. stacking faults). Given the high sensitivity of the van der Waals structure to a mechanical stress, some distortions might be easily introduced during the sample growth or the setup for the thermal-transport measurements. Indeed, a previous transmission electron microscopy [62] confirms a large population of the stacking faults in *TM*$PS_3$ (*TM* = Mn, Ni, Fe), which is pointed out to reduce the cross-plane thermal conductivity of the atomically thin samples. These impurities or distortions might affect the spin interaction parameters as well as the coupling constant between the spins and the lattice, which could alter the field dependence of $\kappa_{xy}$ as pointed out in Ref. [34,49]. A similar effect related to the sample quality is also pointed out in the van der Waals magnet α-$RuCl_3$ (see Ref. [27,63] for example).

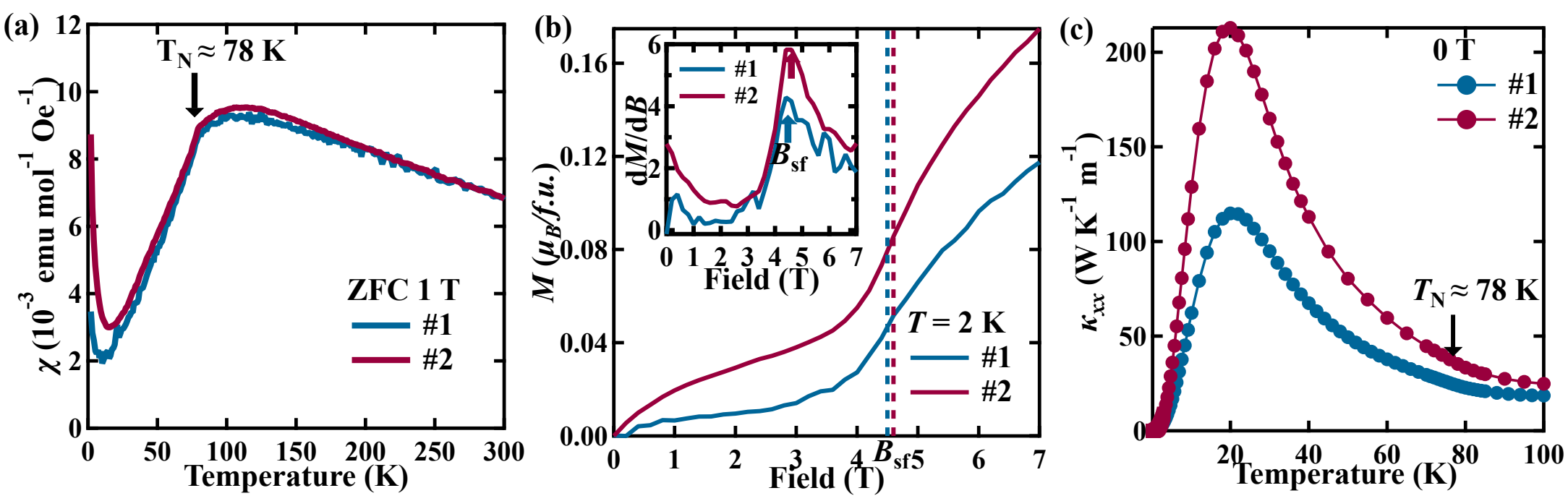

FIG. 10 (a–c) the temperature dependence of the magnetic susceptibility (a), the field dependence of the magnetization (b), and the temperature dependence of the longitudinal thermal conductivity at 0 T (c) of two $MnPS_3$ samples (#1 and #2). The inset in (b) shows the field dependence of the field derivative of the magnetization.

During the preparation of this manuscript, we became aware of a similar study on $MnPS_3$ down to 4.8 K (~$0.02T_{\mathrm{N}}$) that reports a broad sample dependence with a very different magnetic field dependence of $\kappa_{xy}$ with a much larger magnitude from both the previous work [49] and our results, which points out the importance of a magnetic contribution beyond a phonon contribution from a magnon-phonon hybridization [61]. Although the sign reversals in the field dependence of $\kappa_{xy}$ might be reproduced by separating the phonon contribution in the data reported in Ref. [61], the difference of the two-orders of magnitude in $\kappa_{xy}$ from our data and the previous work [49] would be difficult to explain even by considering a magnon-phonon hybridization. This is because, in the formalism of the intrinsic thermal Hall effect (Eq. (1)), $\kappa_{xy}$ is given by a summation of the Berry curvature weighted by a function of Bose factor [3,4]. Given that a finite Berry curvature is caused by a secondary interaction (such as Dzyaloshinskii–Moriya and spin–lattice coupling) and the Nielsen–Ninomiya theorem regarding the Berry curvature in a magnetic insulator, the magnitude of the Berry curvature of each band cannot be extremely large, limiting the order of magnitude of $\kappa_{xy}$ from this intrinsic Berry-curvature expression to $\kappa_{xy}/T \sim k_{\mathrm{B}}^2/(\hbar\, d_c)$ where $d_c$ is the interlayer distance. The magnitude of this $\kappa_{xy}$ is roughly on the order of 1 mW $K^{-1}$ $m^{-1}$ at 10 K, which is in the same order with our results and that from Ref. [49]. In fact, this size of $\kappa_{xy}$ observed in the ferromagnetic skyrmion material [9] and in the antiferromagnetic kagome [20] have been shown to agree with the theoretical calculations by the intrinsic Berry-curvature expression (Eq. (1)) with respect to not only the qualitative temperature dependence of $\kappa_{xy}$ but also their quantitative magnitude.

Therefore, to explain the enhancement of $\kappa_{xy}$ by the two-orders magnitude, one needs to consider a different origin distinct from the intrinsic Berry-curvature expression. One possibility is an extrinsic origin. As known in the anomalous Hall effect in ferromagnetic metals, the extrinsic skew contribution exceeds the intrinsic contribution in a sample with a larger electric conductivity (see Ref. [64] for example). Indeed, it has been demonstrated that a similar extrinsic mechanism appears as a spin thermal Hall effect in magnetic insulators [21] for samples with larger $\kappa_{xx}$. However, given that the magnitude of $\kappa_{xx}$ only differs a factor of 2 or 3 in the data reported in Ref. [61] from our results, it might also be questionable if this can explain the two-orders difference in the magnitude of $\kappa_{xy}$.

**Appendix F: The magnetization data below 2 K in sample #2**

Figure 11 shows the field dependence of $M$ of sample #2 measured at 0.5 K and 1.0 K by using our home-built Faraday-force magnetometer installed in a dilution refrigerator. As shown in Fig. 11, no obvious anomaly is observed in the magnetization curve within our resolution, except the kink at the spin-flop transition at 4.5 T.

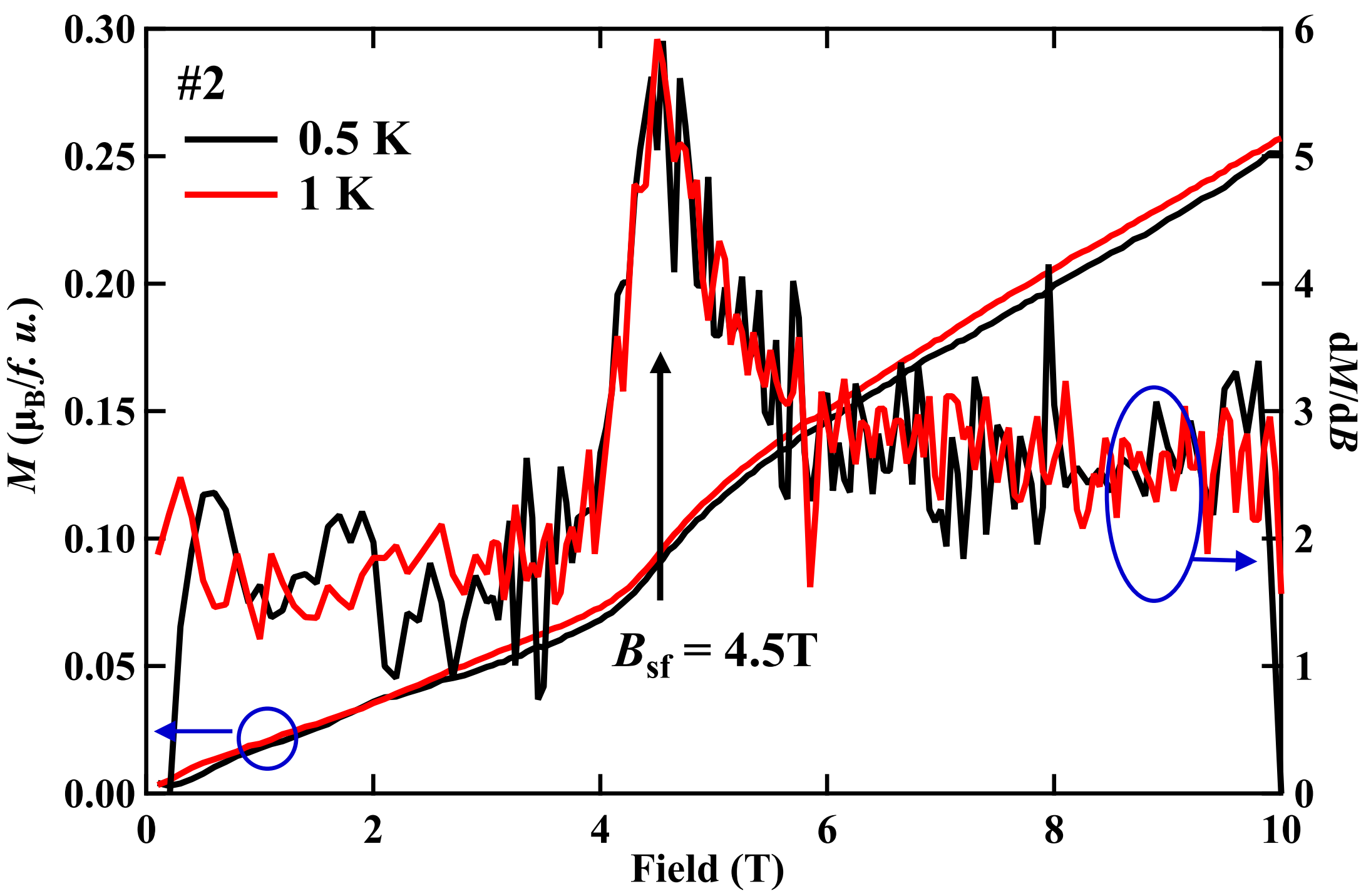

FIG. 11 The magnetic field dependence of the magnetization ($M$) of sample #2 at 0.5 K (black line) and 1 K (red line). The field derivative of the data ($dM/dB$) is also shown on the right axis.